\documentclass[letterpaper, 10 pt, conference]{ieeeconf}  

\IEEEoverridecommandlockouts                              
\usepackage{amsmath} 
\usepackage{amssymb}  
\usepackage{color}
\usepackage{cite}
\usepackage{epsfig} 
\usepackage{graphics} 
\usepackage{graphicx}
\usepackage{ifpdf}
\usepackage{mathptmx} 
\usepackage{subfigure}
\usepackage{times} 
\usepackage{times} 

\newtheorem{theorem}{Theorem}

\newtheorem{corollary}[theorem]{Corollary}

\newtheorem{example}[theorem]{Example}

\newtheorem{lemma}[theorem]{Lemma}

\title{\LARGE \bf
Moment-Based Analysis of Synchronization in Small-World Networks of
Oscillators
}

\author{Victor M. Preciado and Ali Jadbabaie
\thanks{This work was supported by ONR MURI N000140810747, and AFOR's complex networks program.}
\thanks{The authors are with the Department of Electrical and Systems Engineering, University of Pennsylvania, 3451 Walnut Street, 
        {\tt\small \{preciado,jadbabai\}@seas.upenn.edu}}%
}

\begin{document}

\maketitle
\thispagestyle{empty}
\pagestyle{empty}

\begin{abstract}

In this paper, we investigate synchronization in a small-world network of coupled nonlinear oscillators. This network is constructed by introducing random shortcuts in a nearest-neighbors ring. The local stability of the synchronous state is closely related with the support of the eigenvalue distribution of the Laplacian matrix of the network. We introduce, for the first time, analytical expressions for the first three moments of the eigenvalue distribution of the Laplacian matrix as a function of the probability of shortcuts and the connectivity of the underlying nearest-neighbor coupled ring. We apply these expressions to estimate the spectral support of the Laplacian matrix in order to predict synchronization in small-world networks. We verify the efficiency of our predictions with numerical simulations.

\end{abstract}

\section{Introduction}

In recent years, systems of dynamical nodes interconnected through a complex
network have attracted a good deal of attention \cite{S01}. Biological and
chemical networks, neural networks, social and economic networks \cite{Jac08}%
, the power grid, the Internet and the World Wide Web \cite{DM03} are
examples of the wide range of applications that motivate this interest (see
also \cite{New03}, \cite{BLMCH06} and references therein). Several modeling
approaches can be found in the literature \cite{DM03}, \cite{WS98}, \cite%
{BA99}. In this paper, we focus our attention on the so-called small-world phenomenon and a model proposed by Newman and Strogatz to replicate this phenomenon.

Once the network is modeled, one is usually interested in two types of
problems. The first involves \emph{structural properties }of the model. The
second involves the performance of \emph{dynamical processes} run on those
networks. In the latter direction, the performance of random walks \cite%
{Lov93}, Markov processes \cite{Bre01}, gossip algorithms \cite{BGPS06},
consensus in a network of agents \cite{OFM07}, \cite{JLM02}, or
synchronization of oscillators \cite{S03}, \cite{PC98}, are very well
reported in the literature. These dynamical processes are mostly studied in
the traditional context of deterministic networks of relatively small size
and/or regular structure. Even though many noteworthy results have been
achieved for large-scale probabilistic networks \cite{LYCC04}--\cite{SBG07},
there is substantial reliance on numerical simulations.

The \emph{eigenvalue spectrum }of an undirected graph contains a great deal
of information about structural and dynamical properties \cite{C97}. In
particular, we focus our attention on the spectrum of the \emph{%
(combinatorial) Laplacian matrix }uniquely associated with an undirected graph \cite{Big93}.
This spectrum contains useful information about, for example, the number of
spanning trees, or the stability of synchronization of a network of
oscillators. We analyze the low-order moments of the Kirchhoff matrix
spectrum corresponding to small-world networks.

The paper is organized as follows. In Section
II, we review the master stability function approach. In Section III, we
derive closed-form expressions for the low-order moments of the Laplacian
eigenvalue distribution associated with a probabilistic small-world network.
Our expressions are valid for networks of asymptotically large size. Section
IV applies our results to the problem of synchronization of a probabilistic
small-world network of oscillators. The numerical results in this section
corroborate our predictions.

\section{Synchronization of Nonlinear Oscillators}

In this section we review the master-stability-function (MSF) approach,
proposed by Pecora and Carrol in \cite{PC98}, to study local stability of
synchronization in networks of nonlinear oscillators. Using this approach,
we reduce the problem of studying local stability of synchronization to the
algebraic problem of studying the spectral support of the Laplacian matrix
of the network. First, we introduce some needed graph-theoretical background.

\subsection{Spectral Graph Theory Background}

In the case of a network with symmetrical connections, undirected graphs
provide a proper description of the network topology. An undirected graph $G$
consists of a set of $N$ nodes or vertices, denoted by $V=\left\{
v_{1},...,v_{n}\right\} $, and a set of edges $E$, where $E\in V\times V$.
In our case, $\left( v_{i},v_{j}\right) \in E$ implies $\left(
v_{j},v_{i}\right) \in E,$ and this pair corresponds to a single edge with
no direction; the vertices $v_{i}$ and $v_{j}$ are called \emph{adjacent}
vertices (denoted by $v_{i}\sim v_{j}$) and are \emph{incident} to the edge $%
\left( v_{i},v_{j}\right) $. We only consider simple graphs (i.e.,
undirected graphs that have no self-loops, so $v_{i}\neq v_{j}$ for an edge $%
\left( v_{i},v_{j}\right) $, and no more than one edge between any two
different vertices). A \emph{walk }on $G$ of length $k$ from $v_{0}$ to $%
v_{k}$ is an ordered set of vertices $\left( v_{0},v_{1},...,v_{k}\right) $
such that $\left( v_{i},v_{i+1}\right) \in E,$ for $i=0,1,...,k-1$; if $\nu
_{k}=\nu _{0}$ the walk is said to be \emph{closed}.

The \emph{degree} $d_{i}$ of a vertex $v_{i}$ is the number of edges
incident to it. The \emph{degree sequence} of $G$ is the list of degrees,
usually given in non-increasing order. The \emph{clustering coefficient},
introduced in \cite{WS98}, is a measure of the number of triangles in a
given graph, where a triangle is defined by the set of edges $\left\{ \left(
i,j\right) ,\left( j,k\right) ,\left( k,i\right) \right\} $ such that $i\sim
j\sim k\sim i$. Specifically, we define clustering as the total number of
triangles in a graph, $T\left( G\right) ,$ divided by the number of
triangles in a complete (all-to-all) graph with $N$ vertices, i.e., the
coefficient is equal to $T\left( G\right) \left/ \binom{N}{3}\right. .$

It is often convenient to represent graphs via matrices. There are several
choices for such a representation. For example, the \emph{adjacency matrix }%
of an undirected graph $G,$ denoted by $A(G)=[a_{ij}]$, is defined
entry-wise by $a_{ij}=1$ if nodes $i$ and $j$ are adjacent, and $a_{ij}=0$
otherwise. (Note that $a_{ii}=0$ for simple graphs.) Notice also that the
degree $d_{i}$ can be written as $d_{i}=\sum_{j=1}^{N}a_{ij}$. We can
arrange the degrees on the diagonal of a diagonal matrix to yield the \emph{%
degree matrix}, $D=diag\left( d_{i}\right) $. The \emph{Laplacian matrix }%
(also called Kirchhoff matrix, or combinatorial Laplacian matrix) is defined
in terms of the degree and adjacency matrices as $L(G)=D(G)-A(G).$ For
undirected graphs, $L(G)$ is a symmetric positive semidefinite matrix \cite%
{Big93}. Consequently, it has a full set of $N$ real and orthogonal
eigenvectors with real non-negative eigenvalues. Since all rows of $L$ sum
to zero, it always admits a trivial eigenvalue $\lambda _{1}=0$, with
corresponding eigenvector $\mathbf{v}_{1}=\left( 1,1,...,1\right) ^{T}$.

The moments of the Laplacian eigenvalue spectrum are central to our paper.
Denote the eigenvalues of our $N\times N$ symmetric Laplacian matrix $L(G)$
by $0=\lambda _{1}\left( G\right) \leq ...\leq \lambda _{N}\left( G\right) $%
. The \emph{empirical spectral density (ESD)} of $L(G)$ is defined as%
\begin{equation*}
\mu _{G}\left( \lambda \right) =\frac{1}{N}\sum_{i=1}^{N}\delta \left(
\lambda -\lambda _{i}\right) ,
\end{equation*}%
where $\delta (\cdot )$ is the Dirac delta function. The $k$\emph{-}th order
moment of the ESD of $L(G)$ is defined as: 
\begin{equation*}
q_{k}(G)=\frac{1}{N}\sum_{i=1}^{N}\lambda _{i}(G)^{k}
\end{equation*}%
(which is also called the $k$-th \emph{order spectral moment\footnote{%
Given that our interest is in networks of growing size (i.e., number of
nodes $N$), a more explicit notation for $\mu $ and $q_{k}$ would perhaps
have been $\mu ^{\left( N\right) }$ and $q_{k}^{\left( k\right) }$. However,
for notational simplicity, we shall omit reference to $N$ in there and other
quantities in this paper.}}).

In the following subsection, we illustrate how a network of identical
nonlinear oscillators synchronizes whenever the Laplacian spectrum is
contained in a certain region on the real line. This \emph{region of
synchronization }is exclusively defined by the dynamics of each isolated
oscillator and the type of coupling \cite{PC98}, \cite{LC05}. This
simplifies the problem of synchronization to the problem of locating the
Laplacian eigenvalue spectrum.

\subsection{\label{Sync region}Synchronization as a Spectral Graph Problem}

Several techniques have been proposed to analyze the synchronization of
coupled identical oscillators. In \cite{W01}, well-known results in control
theory, such as the passivity criterion, the circle criterion, and a result
on observer design are used to derive synchronization criteria for an array
of identical nonlinear systems. In \cite{SW04}, the authors use contraction
theory to derive sufficient conditions for global synchronization in a
network of nonlinear oscillators. We pay special attention to the
master-stability-function (MSF) approach, \cite{PC98}. This approach
provides us with a criterion for local stability of synchronization based on
the numerical computation of Lyapunov exponents. Even though quite different
in nature, the mentioned techniques emphasize the key role played by the
graph eigenvalue spectrum.

In this paper we consider a time-invariant network of $N$ identical
oscillators, one located at each node, linked with `diffusive' coupling. The
state equations modeling the dynamics of the network are 
\begin{equation}
\mathbf{\dot{x}}_{i}=\mathbf{f}\left( \mathbf{x}_{i}\right) +\gamma
\sum_{j=1}^{N}a_{ij}\Gamma \left( \mathbf{x}_{j}-\mathbf{x}_{i}\right) ,%
\text{ }i=1,...,N  \label{network dynamics}
\end{equation}%
\newline
where $\mathbf{x}_{i}$ represents an $n$-dimensional state vector
corresponding to the $i$-th oscillator. The nonlinear function $\mathbf{f}%
\left( \cdot \right) $ describes the (identical) dynamics of the isolated
nodes. The positive scalar $\gamma $ can be interpreted as a global coupling
strength parameter. The $n\times n$ matrix $\Gamma $ represents how states
in neighboring oscillators couple linearly, and $a_{ij}$ are the entries of
the adjacency matrix. By simple algebraic manipulations, one can write down
Eq. (\ref{network dynamics}) in terms of the Laplacian entries, $L(G)=\left[
l_{ij}\right] $, as 
\begin{equation}
\mathbf{\dot{x}}_{i}=\mathbf{f}\left( \mathbf{x}_{i}\right) -\gamma
\sum_{j=1}^{N}l_{ij}\Gamma \mathbf{x}_{j}\text{, for }i=1,...,N.
\label{Network Dynamics}
\end{equation}

We say that the network of oscillators is at a synchronous equilibrium if $%
\mathbf{x}_{1}(t)=\mathbf{x}_{2}(t)=...=\mathbf{x}_{N}(t)=\mathbf{\phi }%
\left( t\right) $, where $\mathbf{\phi }\left( t\right) $ represents a
solution for $\mathbf{\dot{x}}=\mathbf{f}\left( \mathbf{x}\right) $. In \cite%
{PC98}, the authors studied the local stability of the synchronous
equilibrium. Specifically, they considered a sufficiently small
perturbation, denoted by $\mathbf{\varepsilon }_{i}(t)$, from the
synchronous equilibrium, i.e., 
\begin{equation*}
\mathbf{x}_{i}(t)=\mathbf{\phi }\left( t\right) +\mathbf{\varepsilon }%
_{i}(t).
\end{equation*}%
After appropriate linearization, one can derive the following equations to
approximately describe the evolution of the perturbations: 
\begin{equation}
\mathbf{\dot{\varepsilon}}_{i}=\mathbf{Df}\left( t\right) \,\mathbf{%
\varepsilon }_{i}(t)-\gamma \sum_{j=1}^{n}l_{i,j}\Gamma \,\mathbf{%
\varepsilon }_{j}(t)\text{, for }i=1,...,N.  \label{Variational Equation}
\end{equation}%
where $\mathbf{Df}\left( t\right) $ is the Jacobian of $\mathbf{f}\left(
\cdot \right) $ evaluated along the trajectory $\mathbf{\phi }\left(
t\right) $. This Jacobian is an $n\times n$ matrix with time-variant
entries. Following the methodology introduced in \cite{PC98}, Eq. (\ref%
{Variational Equation}) can be similarity transformed into a set of linear
time-variant (LTV) ODEs of the form: 
\begin{equation}
\mathbf{\dot{\xi}}_{i}\mathbf{=}\left[ \mathbf{Df}\left( t\right) +(\gamma
\lambda _{i}\left( G\right) )~\Gamma \right] \mathbf{\xi }_{i},\text{ for }%
i=1,...,N,  \label{Linear Time-Periodic}
\end{equation}%
where $\{\lambda _{i}\left( G\right) \}_{1\leq i\leq N}$ is the set of
eigenvalues of $L\left( G\right) $. Based on the stability analysis
presented in \cite{PC98}, the network of oscillators in (\ref{network
dynamics}) presents a locally stable synchronous equilibrium if the
corresponding maximal nontrivial Lyapunov exponents of (\ref{Linear
Time-Periodic}) is negative for $i=2,...,N$.

Inspired in Eq. (\ref{Linear Time-Periodic}), Pecora and Carroll studied in 
\cite{PC98} the stability of the following parametric LTV-ODE in the
parameter $\sigma $: 
\begin{equation}
\mathbf{\dot{\xi}=}\left[ \mathbf{Df}\left( t\right) +\sigma \Gamma \right] 
\mathbf{\xi ,}  \label{LTP MSF}
\end{equation}%
where $\mathbf{Df}\left( t\right) $ is the linear time-variant Jacobian in
Eq. (\ref{Variational Equation}). The master stability function (MSF),
denoted by $F\left( \sigma \right) $, is defined as the value of the maximal
nontrivial Lyapunov exponent of (\ref{LTP MSF}) as a function of $\sigma $.
Note that $F\left( \sigma \right) $ depends exclusively on $\mathbf{f}\left(
\cdot \right) $ and $\Gamma $, and is independent of the coupling topology,
i.e., independent of $L\left( G\right) $. The region of synchronization is,
therefore, defined by the range of $\sigma >0$ for which $F\left( \sigma
\right) <0$. For a broad class of systems, the MSF is negative in the
interval $\sigma \in \left[ 0,\sigma _{\max }\right] \equiv S$ (although
more generic stability sets are also possible, we assume, for simplicity,
this is the case in subsequent derivations). In order to achieve
synchronization, the set of scaled nontrivial Laplacian eigenvalues, $%
\{\gamma \lambda _{i}\}_{2\leq i\leq N}$, must be located inside the region
of synchronization, $S$. This condition is equivalent to: $\gamma \lambda
_{2}>0$ and $\gamma \lambda _{N}<\sigma _{\max }$.

We illustrate how to use of the above methodology in the following example:

\bigskip

\begin{example}
\label{Rossler example}Study the stability of synchronization of a ring of 6
coupled R\"{o}ssler oscillators \cite{MMZ04}. The dynamics of each
oscillator is described by the following system of three nonlinear
differential equations: 
\begin{eqnarray*}
\dot{x}_{i} &=&-\left( y_{i}+z_{i}\right) , \\
\dot{y}_{i} &=&x_{i}+a\,y_{i}, \\
\dot{z}_{i} &=&b+z_{i}\left( x_{i}-c\right) .
\end{eqnarray*}
\end{example}

The adjacency entries, $a_{ij}$, of a ring graph of six nodes are $a_{i,j}=1$
if $j\in \{(i+1)\text{ mod }6,(i-1)\text{ mod }6\}$, for $i=1,2,...,6$, and $%
a_{ij}=0$ otherwise. The dynamics of this ring of oscillators are defined
by: 
\begin{equation}
\left[ 
\begin{array}{c}
\dot{x}_{i} \\ 
\dot{y}_{i} \\ 
\dot{z}_{i}%
\end{array}%
\right] =\left[ 
\begin{array}{c}
-\left( y_{i}+z_{i}\right) \\ 
x_{i}+a\,y_{i} \\ 
b+z_{i}\left( x_{i}-c\right)%
\end{array}%
\right] +\gamma \sum_{j\in R\left( i\right) }\left[ 
\begin{array}{c}
x_{j}-x_{i} \\ 
0 \\ 
0%
\end{array}%
\right]  \label{Network of Rosslers}
\end{equation}%
where \ we have chosen to connect the oscillators through their $x_{i}$
states exclusively. Our choice is reflected in the structure of the $3\times
3$ matrix, $\Gamma $, inside the summation in Eqn. (\ref{Network of Rosslers}%
).

\bigskip

Numerical simulations of an isolated R\"{o}ssler oscillator unveil the
existence of a periodic trajectory with period $T=5.749$ when the parameters
in Eqn. (\ref{Network of Rosslers}) take the values $a=0.2,\,b=0.2,$ and $%
c=2.5$ (see Fig. 1). We denote this periodic trajectory by $\mathbf{\phi }%
\left( t\right) =\left[ \phi _{x}\left( t\right) ,\phi _{y}\left( t\right)
,\phi _{z}\left( t\right) \right] $. In our specific case, the LTP
differential equation (\ref{LTP MSF}) takes the following form: 
\begin{equation}
\mathbf{\dot{\xi}}=\left( \left[ 
\begin{array}{ccc}
0 & -1 & -1 \\ 
1 & a & 0 \\ 
\phi _{z}\left( t\right)  & 0 & c%
\end{array}%
\right] +\sigma \left[ 
\begin{array}{ccc}
1 & 0 & 0 \\ 
0 & 0 & 0 \\ 
0 & 0 & 0%
\end{array}%
\right] \right) \mathbf{\xi ,}  \label{LTP ODE Rossler}
\end{equation}%
where the leftmost matrix in the above equation represents the Jacobian of
the isolated R\"{o}ssler evaluated along the periodic trajectory $\mathbf{%
\phi }\left( t\right) ,$ and the rightmost matrix represents $\Gamma $.

\begin{figure}
 \centering
 \includegraphics[width=0.9\linewidth]{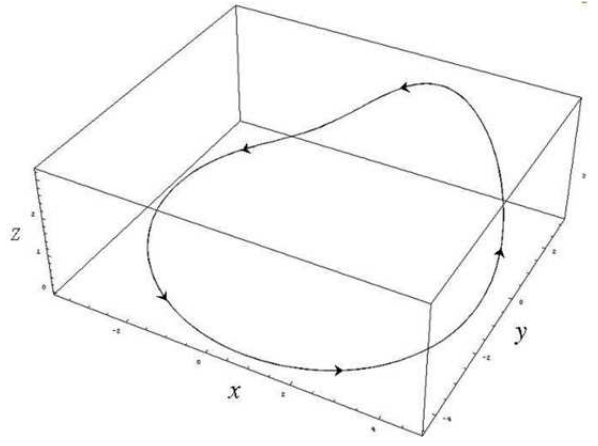}
 \caption{Periodic trajectory with period $T=5.749$ in a R\"{o}ssler oscillator when the parameters
in Eqn. (\ref{Network of Rosslers}) take the values $a=0.2,\,b=0.2,$ and $%
c=2.5$. }
\end{figure}

\begin{figure}
 \centering
 \includegraphics[width=1.0\linewidth]{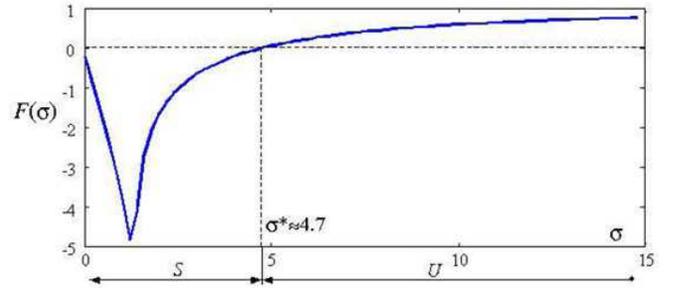}
 \caption{Numerical values of the maximum Floquet exponent of
Eqn. (\ref{LTP ODE Rossler}) for $\sigma \in \left[ 0,15\right] $,
discretizing at intervals of length $0.2$.}
\end{figure}

In Fig. 2, we plot the numerical values of the maximum Floquet exponent of
Eqn. (\ref{LTP ODE Rossler}) for $\sigma \in \left[ 0,15\right] $,
discretizing at intervals of length $0.2$. This plot shows the range in
which the maximal Floquet exponent is negative. This range of stability is $%
S=\left( 0,\sigma ^{\ast }\right) $, for $\sigma ^{\ast }\approx 4.7$. The
MSF criterion introduced in \cite{PC98} states that the synchronous
equilibrium is locally stable if the set of values $\{\gamma ~\lambda
_{i}\left( G\right) \}_{i=2,...,n}$ lies inside the stability range, $S$.
For the case of a 6-ring configuration, the eigenvalues of $L\left( G\right) 
$ are $\left\{ {0,1,1,3,3,4}\right\} $, so the set $\{\gamma \lambda
_{i}\}_{i=2,...,n}$ is $\left\{ \gamma ,\gamma ,3\gamma ,3\gamma ,4\gamma
\right\} .$ Therefore, we achieve stability for $\gamma \in (0,\sigma ^{\ast
}/\lambda _{n}\left( G\right) )$, where in our case $\sigma ^{\ast }/\lambda
_{n}\left( G\right) \approx 1.175$.

We now illustrate this result with several numerical simulations. First, we
plot in Fig. 3.a the temporal evolution of the $x_{i}$ states of the 6-ring
when $\gamma =1.0$. Observe how, since $\gamma \in \left( 0,1.175\right) $,
we achieve asymptotic synchronization. On the other hand, if we choose $%
\gamma =1.3\notin \left( 0,1.175\right) $, the time evolution of the set of
oscillators does not converge to a common trajectory (see Fig. 3.b);
instead, the even and odd nodes settle into two different trajectories.

\begin{figure*}
 \centering
 \includegraphics[width=0.95\linewidth]{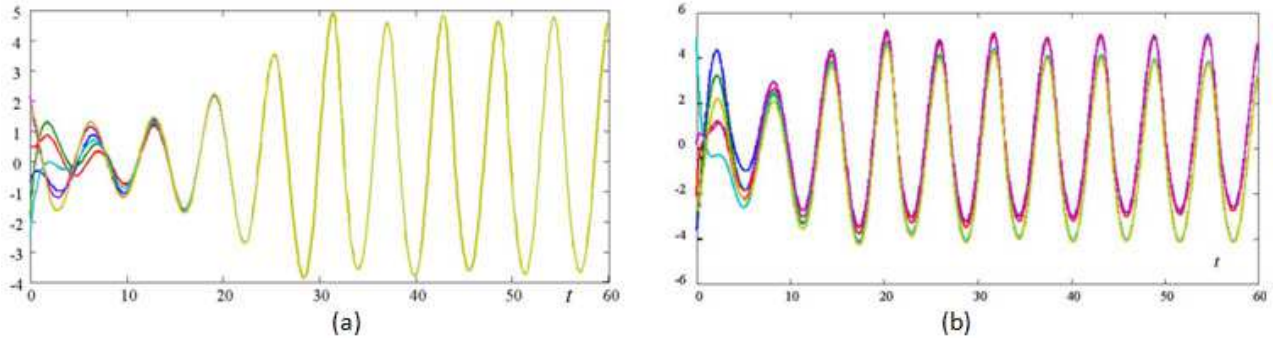}
 \caption{In Fig. a, we plot the temporal evolution of the $x_{i}$ states of the 6-ring
when $\gamma =1.0$. In Fig.b, we plot the time evolution of the set of
oscillators for $\gamma =1.3\notin \left( 0,1.175\right)$.}
\end{figure*}

In the next subsection, we propose an approach to estimating the support of
the eigenvalue distribution of large-scale probabilistic networks from
low-order spectral moments. This allows us to predict synchronization in a
large-scale Chung-Lu network.

\section{Spectral Analysis of Small-World Networks}

In this section we study the Laplacian eigenvalue spectrum of a variant of
Watts-Strogatz small-world network \cite{WS98}. After describing the model,
we use algebraic graph theory to compute explicit expressions for the
Laplacian moments of a small-world network as a function of its parameters.
Our derivations are based on a probabilistic analysis of the expected
spectral moments of the Laplacian for asymptotically large small-world
networks.

\subsection{\label{SW model}Small-World Probabilistic Model}

We consider a one-dimensional lattice of $N$ vertices, $\left\{
v_{1},...,v_{N}\right\} $, with periodic boundary conditions, i.e., on a
ring, and connect each vertex $v_{i}$ to its $2k$ closest neighbors, i.e., $%
v_{i}$ is connected to the set of nodes $\left\{ v_{j}:j\in \left[ \left(
i-k\right) \text{mod}N,\left( i+k\right) \text{mod}N\right] \right\} $.
Then, instead of rewiring a fraction of the edges in the regular lattice as
proposed by Watts and Strogatz \cite{WS98}, we add some random `shortcuts'
to the one-dimensional lattice. These shortcuts are added by independently
assigning edges between each pair of nodes $\left( i,j\right) ,$ $1\leq
i<j\leq N$ with probability $p$. The resulting small-world graph is
intermediate between a regular lattice (achieved for $p=0$) and a classical
random graph (achieved for $p=1$). In general, \emph{small-world networks}
share properties with both the regular grid and the classical random graph
for $0<p<1$. In particular, they show the following apparently contradictory
features:

\emph{(i)} most nodes are not neighbors of one another (such as in a regular
grid), and

\emph{(ii)} most nodes can be reached from every other node by a small
number of steps (such as in a random graph).

An interesting property observed in this model was the following: for small
probability of rewiring, $p\ll 1$, the number of triangles in the network is
nearly the same as that of the regular lattice, but the average
shortest-path length is close to that of classical random graphs. In the
rest of the paper we assume we are in the range of $p$ in which this
property holds, in particular, we will prescribe $p$ to be \thinspace $r/N$,
for a given parameter $r$.

In the coming sections we shall study spectral properties of the Laplacian
matrix associated to the above small-world model. In our derivations we will
need the probabilistic distribution for the degrees. It is well known that,
for asymptotically large graphs, the degree distribution of a classical
random graph with average degree $r$ is a Poisson distribution with rate $r$%
. Hence, the degree distribution of the above small-world network is%
\begin{equation}
\Pr (d_{i}=d)=\left\{ 
\begin{array}{ll}
0, & \text{for }d<2k, \\ 
\frac{r^{d-2k}e^{-r}}{(d-2k)!}, & \text{for }d\geq 2k,%
\end{array}%
\right.  \label{Shifted Poisson}
\end{equation}%
which corresponds to a Poisson with parameter $r$ `shifted' $2k$ units. The
Poisson distribution is shifted to take into account the degree of the
regular $2k$-neighbors ring superposed to the random shortcuts.

Furthermore, it is well known that the clustering coefficient (or,
equivalently, the number of triangles) of the regular $2k$-neighbors rings
is very lightly perturbed by the addition of random shortcuts for $p=r/N$.
In particular, one can prove the following result:%
\begin{equation}
\mathbb{E}[T]=(1+o(1))\,\frac{1}{3}N\binom{2k}{2},  \label{Triangles SW}
\end{equation}%
where the dominant term, $\frac{1}{3}N\binom{2k}{2}$, corresponds to the
exact number of triangles in a $2k$-neighbors ring with $N$ nodes.

In the following section, we shall derive explicit expressions for the first
low-order spectral moments of the Laplacian matrix associated with the
small-world model herein described. Even though our analysis is far from
complete, in that only low-order moments are provided, valuable information
regarding spectral properties can be retrieved from our results.

\subsection{Algebraic Analysis of Spectral Moments}

In this section we deduce closed-form expressions for the first three
moments of the Laplacian spectrum of any simple graph $G$. First, we express
the spectral moments as a trace using the following identity: 
\begin{equation}
q_{k}\left( G\right) =\frac{1}{N}\sum_{i=1}^{N}\lambda _{i}\left( G\right)
^{k}=\frac{1}{N}\,\text{tr}\,L\left( G\right) ^{k}.
\end{equation}%
This identity is derived from the fact that trace is conserved under
diagonalization (in general, under any similarity transformation). In the
case of the first spectral moment, we obtain 
\begin{equation*}
q_{1}=\frac{1}{N}\text{tr}\left( D-A\right) =\frac{1}{N}\sum_{i=1}^{N}d_{i}.
\end{equation*}%
where $\overline{d}$ is the average degree of the graph. For analytical and
numerical reasons, we define the normalized Kirchhoff moment as 
\begin{equation}
\overline{q}_{k}=\frac{1}{N}\sum_{i=1}^{N}\left( \lambda _{i}/\overline{d}%
\right) ^{k}=\frac{1}{N\,\overline{d}^{k}}\,\text{tr}\left( D-A\right) ^{k}.
\label{normalized moment}
\end{equation}

The fact that $D$ and $A$ do not commute forecloses the possibility of using
Newton's binomial expansion on $\left( D-A\right) ^{k}$. On the other hand,
the trace operator allows us to cyclically permute multiplicative chains of
matrices. For example, tr$\,\left( AAD\right) =$tr$\,\left( ADA\right) =$tr$%
\,\left( DAA\right) $. Thus, for words of length $k\leq 3$, one can
cyclically arrange all binary words in the expansion of (\ref{normalized
moment}) into the standard binomial expression: 
\begin{equation}
\overline{q}_{k}=\sum_{\alpha =0}^{k}\binom{k}{\alpha }\frac{(-1)^{\alpha }}{%
\overline{d}^{k}N}\text{tr\thinspace }\left( A^{\alpha }D^{k-\alpha }\right) 
\text{,\quad \quad for }k\leq 3.
\end{equation}
Also, we can make use of the identity tr$\left( A^{\alpha }D^{k-\alpha
}\right) =\sum_{i=1}^{N}\left( A^{\alpha }\right) _{ii}d_{i}^{k-\alpha }$ to
write 
\begin{equation}
\overline{q}_{k}=\sum_{\alpha =0}^{k}\sum_{i=1}^{N}\binom{k}{\alpha }\frac{%
(-1)^{\alpha }}{\overline{d}^{k}N}d_{i}^{k-\alpha }\left( A^{\alpha }\right)
_{ii}\text{,\quad \quad for }k\leq 3.  \label{newton-like}
\end{equation}
Note that this expression is not valid for $k\geq 4$. For example, for $k=4$%
, we have that tr$\left( AADD\right) \neq $tr$\left( DADA\right) .$

We now analyze each summand in expression (\ref{newton-like}) from a
graph-theoretical point of view. Specifically, we find a closed-form
solution for each term tr$\left( A^{i}D^{j}\right) $, for all pairs $1\leq
i+j\leq 3$, as a function of the degree sequence and the number of triangles
in the network. In our analysis, we make use of the following results from 
\cite{Big93}:

\bigskip

\begin{lemma}
The number of closed walks of length $\alpha $ in a graph $G$, joining node $%
i$ to itself, is given by the $i$-th diagonal entry of the matrix $A^{\alpha
}$.
\end{lemma}

\bigskip

\begin{corollary}
Let $G$ be a simple graph. Denote by $t_{i}$ the number of triangles
touching node $i$. Then, 
\begin{equation}
\left( A\right) _{ii}=0,\text{ }\left( A^{2}\right) _{ii}=d_{i},\text{ and }%
\left( A^{3}\right) _{ii}=2\,t_{i}.  \label{algebraic graph}
\end{equation}
\end{corollary}

\bigskip

After substituting (\ref{algebraic graph}) into (\ref{newton-like}), and
straightforward algebraic simplifications, we obtain the following exact
expression for the low-order normalized spectral moments of a given
Kirchhoff matrix $K$:%
\begin{equation}
\overline{q}_{k}=\left\{ 
\begin{array}{lc}
1, & \text{for }k=1, \\ 
\frac{1}{N\,\overline{d}^{2}}\left(
\sum_{i=1}^{N}d_{i}^{2}+\sum_{i=1}^{N}d_{i}\right) , & \text{for }k=2, \\ 
\frac{1}{N\,\overline{d}^{3}}\left[ \left(
\sum_{i=1}^{N}\,d_{i}^{3}+3\sum_{i=1}^{N}d_{i}^{2}\right) -6\,T\right] , & 
\text{for }k=3,%
\end{array}%
\right.  \label{low order moments}
\end{equation}%
where $T=\frac{1}{3}\sum_{i=1}^{N}t_{i}$ is the total number of triangles%
\footnote{%
A triangle is defined by a set of (undirected) edges $\left\{ \left(
i,j\right) ,\left( j,k\right) ,\left( k,i\right) \right\} $ such that $i\sim
j\sim k\sim i$.} in the network.

It is worth noting how our spectral results are written in terms of two
widely reported measurements, \cite{New03}: the \emph{degree sequence} and
the \emph{clustering coefficient }(which provides us with the total number
of triangles.) This allows us to compute low-order spectral moments of many
real-world networks without performing an explicit eigenvalue decomposition.

\bigskip

\subsection{\label{SW moments}Probabilistic Analysis of Spectral Moments}

In this section, we use Eq. (\ref{low order moments}) to compute the first
three expected Laplacian moments of the small-world model under
consideration. The expected moments can be computed if we had explicit
expressions for the moments of the degrees, $\mathbb{E}[d_{i}]$, $\mathbb{E}%
[d_{i}^{2}]$, \ and $\mathbb{E}[d_{i}^{3}]$, and the expected number of
triangles, $\mathbb{E}[T]$. Since we know the degree distribution (\ref%
{Shifted Poisson}) for this model, the moments of the degrees can be
computed to be:%
\begin{eqnarray}
\mathbb{E}[d_{i}] &=&r+2k,  \label{Degrees SW} \\
\mathbb{E}[d_{i}^{2}] &=&r^{2}+\left( 1+4k\right) \,r+4k^{2},  \notag \\
\mathbb{E}[d_{i}^{3}] &=&r^{3}+15\left( 3+6k\right) \,r^{2}+\left(
1+6k+12k^{2}\right) ~r-8k^{3}.  \notag
\end{eqnarray}%
We can therefore substitute the expressions (\ref{Triangles SW}) and (\ref%
{Degrees SW}) in Eq. (\ref{low order moments}) in order to derive the
following expressions for the (non-normalized) expected Laplacian moments
for $N\rightarrow \infty $:%
\begin{eqnarray}
\mathbb{E}[q_{1}] &\rightarrow &\,r+2k,  \label{Moments SW} \\
\mathbb{E}[q_{2}] &\rightarrow &\,r^{2}+\left( 4k+2\right) \,r+4k^{2}+2k, 
\notag \\
\mathbb{E}[q_{3}] &\rightarrow &r^{3}+\left( 6k+6\right) \,r^{2}+\left(
12k^{2}+18k+4\right) \,r  \notag \\
&&+8k^{2}+8k^{3}+2k.  \notag
\end{eqnarray}

In the following table we compare the numerical values of the Laplacian
moments corresponding to one random realization of the model under
consideration with the analytical predictions in (\ref{Moments SW}). In
particular, we compute the moments for a network of $N=512$ nodes with
parameters $p=r/N=4/N$ and $k=3.$ It is important to point out that the indicated numerical values are obtained for one realization only, with no
benefit from averaging.%
\begin{equation*}
\begin{tabular}{llll}
\hline
Moment order & 1$^{st}$ & 2$^{nd}$ & 3$^{rd}$ \\ \hline
Numerical realization & 10.14 & 116.96 & 1,467.6 \\ 
Analytical expectations & 10 & 114 & 1,431 \\ 
Relative error & 1.38\% & 2.53\% & 2.49\% \\ \hline
\end{tabular}%
\end{equation*}

In the next subsection, we use an approach introduced in \cite{P07}\ to
estimate the support of the eigenvalue distribution using the first three
spectral moments. In coming sections, we shall use this technique to predict
whether the Laplacian spectrum lies in the region of synchronization.

\subsection{\label{Triangular Fitting}Piecewise-Linear Reconstruction of the
Laplacian Spectrum}

Our approach, described more fully in \cite{P07}, approximates the spectral
distribution with a triangular function that exactly preserves the first
three moments. We define a triangular distribution $t\left( \lambda \right) $
based on a set of abscissae $x_{1}\leq x_{2}\leq x_{3}$ as%
\begin{equation*}
t\left( \lambda \right) :=\left\{ 
\begin{array}{ll}
\frac{h}{x_{2}-x_{1}}\left( \lambda -x_{1}\right) , & \text{for }\lambda \in %
\left[ x_{1},x_{2}\right) , \\ 
\frac{h}{\left( x_{2}-x_{3}\right) }\left( \lambda -x_{3}\right) , & \text{%
for }\lambda \in \left[ x_{2},x_{3}\right] , \\ 
0, & \text{otherwise.}%
\end{array}%
\right.
\end{equation*}%
where $h=2/\left( x_{3}-x_{1}\right) $. The first three moments of this
distribution, as a function of the abscissae, are given by%
\begin{eqnarray}
M_{1} &=&\frac{1}{3}\left( x_{1}+x_{2}+x_{3}\right) ,  \label{Sym Moments} \\
M_{2} &=&\frac{1}{6}\left(
x_{1}^{2}+x_{2}^{2}+x_{3}^{2}+x_{1}x_{2}+x_{1}x_{3}+x_{2}x_{3}\right) , 
\notag \\
M_{3} &=&\frac{1}{10}\left(
x_{1}^{3}+x_{1}^{2}x_{2}+x_{1}^{2}x_{3}+x_{2}^{3}+x_{2}^{2}x_{1}\right. 
\notag \\
&&+\left.
x_{2}^{2}x_{3}+x_{3}^{3}+x_{3}^{2}x_{1}+x_{3}^{2}x_{2}+x_{1}x_{2}x_{3}%
\right) .  \notag
\end{eqnarray}%
Our task is to find the set of values $\left\{ x_{1},x_{2},x_{3}\right\} $
in order to fit a given set of moments $\left\{ M_{1},M_{2},M_{3}\right\} $.
The resulting system of algebraic equations is amenable to analysis, based
on the observation that the moments are symmetric polynomials\footnote{%
A symmetric polynomial on variables $\left( x_{1},x_{2},x_{3}\right) $ is a
polynomial that is unchanged under any permutation of its variables.}.
Following the methodology in \cite{P07}, we can find the abscissae $\left\{
x_{1},x_{2},x_{3}\right\} $ as roots of the polynomial: 
\begin{equation}
x^{3}-\Pi _{1}x^{2}+\Pi _{2}x-\Pi _{3}=0,  \label{Polynomial Roots}
\end{equation}%
where 
\begin{eqnarray}
\Pi _{1} &=&3\,M_{1},  \label{ESD from moments} \\
\Pi _{2} &=&9\,M_{1}^{2}-6\,M_{2},  \notag \\
\Pi _{3} &=&27\,M_{1}^{3}-36\,M_{1}M_{2}+10\,M_{3}.  \notag
\end{eqnarray}

The following example illustrates how this technique provides a reasonable
estimation of the Laplacian spectrum for small-world Networks.

\begin{example}
\label{Triangle example}Estimate the spectral support of the small-world
model described in Subsection \ref{SW model} for parameters $N=512,$ $p=4/N$
and $k=3$. In subsection \ref{SW moments} we computed the expected spectral
moments of this particular network to be $\left\{
M_{1}=10,M_{2}=114,M_{3}=1,431\right\} $. Thus, we apply the above technique
with these particular values of the moments to compute the following set of
abscissae for the triangular reconstruction $\left\{
x_{1}=1.577,x_{2}=8.662,x_{3}=19.76\right\} $. In Fig. 4 we compare the
triangular function that fits the expected spectral moments with the
histogram of the eigenvalues of one random realization of the Laplacian
matrix. We also observe that any random realization of the eigenvalue
histograms of the Laplacian is remarkably close to each other. Although a
complete proof of this phenomenon is beyond the scope of this paper, one can
easily proof using the law of large numbers that the distribution of
spectral moments in (\ref{low order moments}) concentrate around their mean
values.
\end{example}

\begin{figure}
 \centering
 \includegraphics[width=0.8\linewidth]{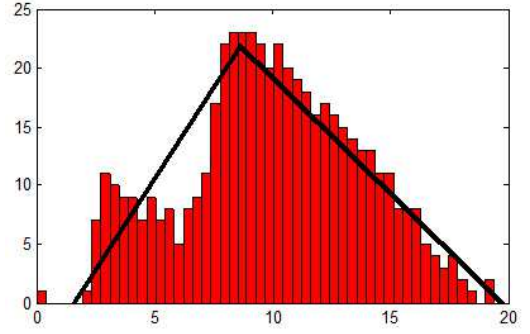}
 \caption{Comparison between the histogram of the eigenvalues of one random realization of the Laplacian
matrix of a small-world model with parameters $N=512,$ $p=4/N$ and $k=3$, and the
triangular function that fits the expected spectral moments.}
\end{figure}

We observe that the above estimation is valid for a large range in the
values of the parameters. For example, in Fig. 5, we compare the values of
the triangular abscissae $x_{1}$ and $x_{3}$ with the extreme points of the
Laplacian spectral support, $\lambda _{2}$ and $\lambda _{n}$, for a
small-world network with $N=512$ nodes, $k=3$, and $p$ in the range of
values $\left[ 1/N:0.01/N:10/N\right] .$ It is important to point out that,
in this case too, the numerical values for the eigenvalues are obtained for
one realization only, with no benefit from averaging. In the next section,
we propose a methodology which uses results presented in previous sections
to predict the local stability of the synchronous state in a small-world
network of oscillators.

\begin{figure}
 \centering
 \includegraphics[width=0.8\linewidth]{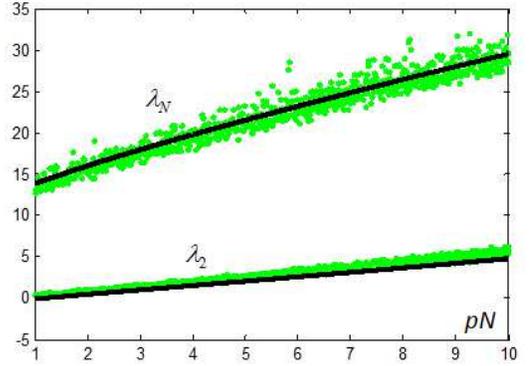}
 \caption{Comparison between the values of
the triangular abscissae $x_{1}$ and $x_{3}$ with the extreme points of the
Laplacian spectral support, $\lambda _{2}$ and $\lambda _{n}$, for a
small-world network with $N=512$ nodes, $k=3$, and $p$ in the range of
values $\left[ 1/N:0.01/N:10/N\right] .$}
\end{figure}

\section{Analytical Estimation of Synchronization}

In this section we use the expressions in (\ref{Moments SW}) and the
triangular reconstruction in the above subsection to predict synchronization
in a large small-world network of coupled nonlinear oscillators.
Specifically, we study a network of coupled R\"{o}ssler oscillators, as
those in Example \ref{Rossler example}. We build our prediction based on the
following steps:

\begin{enumerate}
\item Determine the \emph{region of synchronization} following the technique
presented in Subsection \ref{Sync region}. As illustrated in Example \ref%
{Rossler example}, the \emph{set of scaled eigenvalues} $\{\gamma \lambda
_{i}^{\left( K\right) }\}_{i=2,...,N}$ must lie in a certain region of
stability, $S$, to achieve synchronization (in our example $S=\left(
0,\sigma ^{\ast }\approx 4.7\right) $).

\item Compute the \emph{expected spectral moments} of the Laplacian
eigenvalue spectrum for a given set of parameters using the set of Eqns. in (%
\ref{Moments SW}).

\item Estimate the \emph{support of the Laplacian eigenvalue spectrum,} $%
\{\lambda _{i}^{\left( K\right) }\}_{i=2,...,N}$,\ using the methodology
presented in Subsection \ref{Triangular Fitting}. From Example \ref{Triangle
example}, we have that $s_{l}=1.57$ and $s_{u}=19.76$ are good estimates of
the lower and upper extremes of the spectral support, respectively.

\item Compare the region of stability in \textbf{Step 1} with the estimation
of the spectral support in \textbf{Step 3}, i.e., $\left( 1.57~\gamma
,19.76~\gamma \right) $.

\end{enumerate}

Following the above steps, one can easily verify that our estimated spectral
support, $\left( 1.57~\gamma ,19.76~\gamma \right) $, lies inside the region
of stability, $\left( 0,\sigma ^{\ast }\approx 4.7\right) $, for $0<\gamma
<4.7/19.76\approx 0.238$. Therefore, the small-world network of $512$
coupled R\"{o}ssler oscillators is predicted to synchronize whenever the
global coupling strength satisfies $\gamma \in \left( 0,0.238\right) $.

\subsection{Numerical Results}

\begin{figure*}
 \centering
 \includegraphics[width=0.9\linewidth]{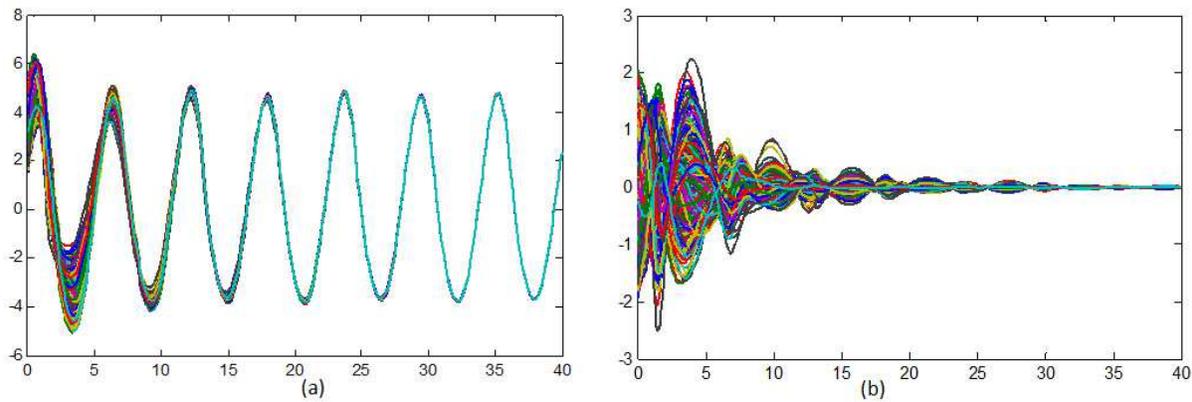}
 \caption{We plot the dynamics of the $x$-states for 512 R\"{o}ssler oscillators (as the one
described in Example \ref{Rossler example}) interconnected through the
Small--World network with $p=4/N$ and $k=3$, in Fig.a. In Fig.b, we observe a clear
exponential convergence of the errors towards zero.}
\end{figure*}

In this section we present numerical simulations supporting our conclusions.
We consider a set of identical $512$ R\"{o}ssler oscillators (as the one
described in Example \ref{Rossler example}) interconnected through the
Small--World network defined in Example \label{Triangle example copy(1)} ($%
p=4/N$ and $k=3$). Using the methodology proposed above, we have predicted
that the synchronous state of this system is locally stable if the coupling
parameter $\gamma $ lies in the interval $\left( 0,0.238\right) $. We run
several simulations with the dynamics of the oscillators presenting
different values of the global coupling strength $\gamma $. For each
coupling strength, we present two plots: \emph{(i)} the evolution of
the\thinspace $512$ $x$-states of the R\"{o}ssler oscillators in the time
interval $0\leq t\leq 40$, and \emph{(ii)} the evolution of $x_{i}\left(
t\right) -\bar{x}\left( t\right) $ for all $i$, where $\bar{x}\left(
t\right) =\frac{1}{N}\sum_{i}x_{i}\left( t\right) $. Since our stability
results are local, we have to carefully choose the initial states for the
network of oscillators. For our particular choice of parameters, the
(isolated) R\"{o}ssler oscillator presents a stable limit cycle. For our
simulations, we have chosen as initial condition for each oscillator in the
network a randomly perturbed version of a particular point of this stable
limit cycle. This particular point is $\mathbf{s}_{0}=\left(
3.5119,-3.5332,0.2006\right) $. We have chosen the perturbed initial state
for the $i$-th oscillator to be $\mathbf{s}_{0}+\mathbf{e}_{i}$, where $%
\mathbf{e}_{i}$ is a uniformly distributed random variable in the $3$%
-dimensional cube $\left[ -2,2\right] ^{3}$, and $\mathbf{e}_{i}$ is
independent of $\mathbf{e}_{j}$ for $i\neq j$.I

In our first simulation, we use a coupling strength $\gamma =0.1\in \left(
0,0.238\right) $; thus, we predict the synchronous state to be locally
stable. Fig. 6 (a) and (b) represents the dynamics $x$-states for the 512
oscillators in the small-world network. In this case, we observe a clear
exponential convergence of the errors to zero. In the second simulation, we
choose $\gamma =0.3\notin \left( 0,0.238\right) $; thus, we predict the
synchronous state to be unstable. In fact, we observe in Figs. 7.a and
7.b how synchronization is clearly not achieved.

\begin{figure*}
 \centering
 \includegraphics[width=0.9\linewidth]{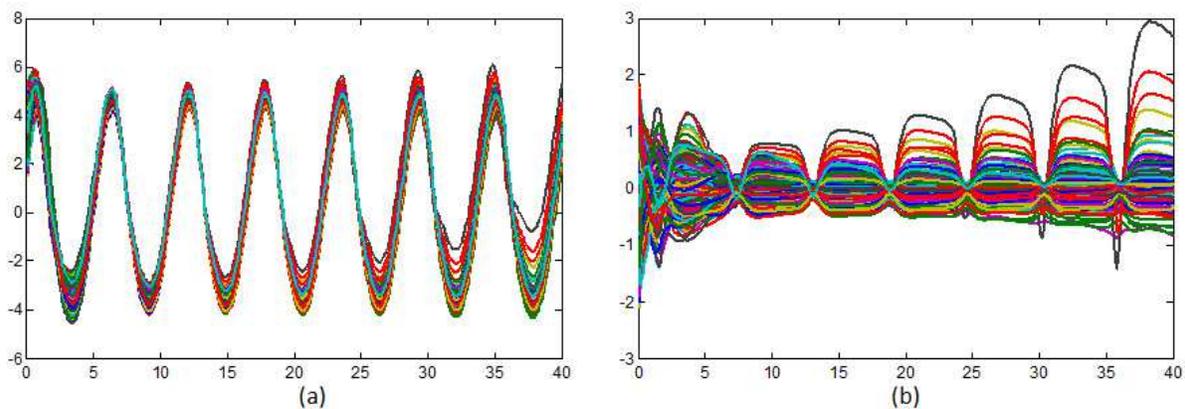}
 \caption{We plot the dynamics of the $x$-states for the 512 R\"{o}ssler oscillators  for $\gamma =0.3\notin \left( 0,0.238\right) $ in Fig.a. We observe in Fig.b. how the errors do not converge to zero.}
\end{figure*}

\section{Conclusions and Future Research}

In this paper, we have studied the eigenvalue distribution of the Laplacian
matrix of a large-scale small-world networks. We have focused our attention
on the low-order moments of the spectral distribution. We have derived
explicit expressions of these moments as functions of the parameters in the
small-world model. We have then applied our results to the problem of
synchronization of a network of nonlinear oscillators. Using our
expressions, we have studied the local stability of the synchronous state in
a large-scale small-world network of oscillators. Our approach is based on
performing a triangular reconstruction matching the first three moments of
the unknown spectral measure. Our numerical results match our predictions
with high accuracy. Several questions remain open. The most obvious
extension would be to derive expressions for higher-order moments of the
Kirchhoff spectrum. A more detailed reconstruction of the spectral measure
can be done based on more moments.

\section{ACKNOWLEDGMENTS}

The first author gratefully acknowledges George C. Verghese and Vincent Blondel for their comments and suggestions on this work.


\end{document}